\documentclass[aps,prl,amsmath,amssymb,twocolumn,showpacs,floatfix]{revtex4-1}
\usepackage{bm}
\usepackage{epsfig}
\usepackage[usenames]{color}
\usepackage{soul}
\usepackage{graphicx}
\usepackage[hypertex]{hyperref}
\usepackage{srctex}
\usepackage{pst-plot}
\usepackage{pstricks}
\renewcommand{\paragraph}[1]{\textit{#1.---} } 

\begin{document}

\title{Monte Carlo modeling the phase diagram of magnets with the Dzyaloshinskii - Moriya interaction}

\author{A.~M.~Belemuk}
\affiliation{Institute for High Pressure Physics, Russian Academy of Science, Troitsk 108840, Russia}
\author{S.~M.~Stishov}
\email{sergei@hppi.troitsk.ru}
\affiliation{Institute for High Pressure Physics, Russian Academy of Science, Troitsk 108840, Russia}
\begin{abstract}
We use classical Monte Carlo calculations to model the high-pressure behavior of the phase transition in the helical magnets. We vary values of the exchange interaction constant $J$ and the Dzyaloshinskii-Moriya interaction constant $D$, which is equivalent to changing spin-spin distances, as occurs in real systems under pressure. The system under study is self-similar at $D/J=constant$, and its properties are defined by the single variable $J/T$, where $T$ is temperature. The existence of the first order phase transition critically depends on the ratio $D/J$. A variation of $J$ strongly affects the phase transition temperature and width of the fluctuation region (the "hump") as follows from the system self-similarity. The high-pressure behavior of the spin system depends on the  evolution of the interaction constants $J$ and $D$ on compression. Our calculations are relevant to the high pressure phase diagrams of helical magnets MnSi and Cu$_2$OSeO$_3$.
\end{abstract}

\maketitle

\section{Introduction}

As is well known the phase transition, or critical, temperature $T_{c}$ in most itinerant magnets decreases with pressure and tends to zero where the phase transition becomes quantum \cite{Brando16}. One of the typical examples of this kind is the phase transition in the helical magnet MnSi with the Dzyaloshinskii-Moriya (DM) interaction \cite{Stishov11}. The opposite example is the insulator Cu$_2$OSeO$_3$ with local magnetic moments, whose the phase transition temperature increase with pressure \cite{Sidorov14}.

Generally, magnetic phase transitions are continuous owing to time reversible symmetry. However, at ambient pressure the phase transition in MnSi and Cu$_2$OSeO$_3$ are weak first order, probably induced by helical fluctuations \cite{Bak80, Janoschek13, Zivkovic14}.

The evolution of the phase transition in a model spin system with ferromagnetic interaction was studied using the Monte Carlo technique in Ref. \cite{Belemuk17} by varying the DM term in the Hamiltonian. It was shown that the puzzling "humps" in the specific heat and other properties at the phase transition in MnSi  originate from smearing out of the virtual ferromagnetic second order phase transition by helical fluctuations. These fluctuations finally condense into the helical ordered phase via a first order phase transition, as is indicated by the sharp peak in specific heat \cite{Belemuk17}.
	
In the present paper we use classical Monte Carlo calculations of the heat capacity to model the high-pressure behavior of the phase transition in helical magnets. The Hamiltonian of the problem does not contain a length term (see below). Thus, the spin system under study is governed by three parameters, the interaction exchange constant $J$, the Dzyaloshinskii-Moriya interaction constant $D$, and temperature $T$. At a fixed ratio $D/J$, only $J$ and $T$ left as the variables. They always appear in the partition function as the ratio $J/T$. Because of that the system becomes self-similar with the single dimensionless variable $J/T$, defining all the system properties. 

So, to model the pressure response of the system we vary values of the exchange interaction $J$ and the DM interaction $D$, which is equivalent to changing spin-spin distances as occurs in real systems under pressure. In principal it would be enough to make calculations with a single value of $J$ as a function of temperature, and to receive  the rest data from scaling. However, we have made the calculations with various $J$ for better statistics and reliability.  Results of the calculations show that the existence of the first order phase transition critically depends on the ratio $D/J$.

A variation of $J$ strongly affects the phase transition temperature and width of the fluctuation region (the "hump") as follows from the system self-similarity and corresponding scaling. The phase transition temperature $T_{c}$ is a linear function of the exchange constant $J$ at fixed value of $D/J$. However, the pressure dependence of $T_{c}$ depends mainly on the behavior of the exchange interaction. We demonstrate that the pressure dependence of the phase transition temperature $T_{c}= T_c(P)$ in the itinerant helimagnet MnSi and in the helimagnet with localized spins Cu$_2$OSeO$_3$ can be modeled using the appropriate kinds of evolution of the interaction constants $J$ and $D$.

\section{Model and simulation}

We use the lattice Hamiltonian, \cite{Hamann11, Yu10, Yi09, Milde13, Ambrose13, Buhrandt13} consisting of Heisenberg exchange ($H_J$) and DM ($H_D$) interaction terms for the microscopic description of a chiral helimagnet
\begin{multline}
H_J+ H_D= -J \sum \limits_{i} {\bf S}_i \cdot ({\bf S}_{i+ \hat x}+ {\bf S}_{i+ \hat y}+ {\bf S}_{i+ \hat z})- \\
-D \sum \limits_{i}  \left({\bf S}_i \times {\bf S}_{i+ \hat x} \cdot \hat x+ {\bf S}_i \times {\bf S}_{i+ \hat y} \cdot \hat y+ {\bf S}_i \times {\bf S}_{i+ \hat z} \cdot \hat z \right)
\end{multline}
where $J$ is the ferromagnetic exchange constant and $D$ is the DM interaction constant.
Classical spins of unit length ${\bf S}_i$ are arranged on a simple cubic lattice of size $N= L^3$ spanned by the vectors $\hat x$, $\hat y$ and $\hat z$. The lattice spacing is taken to be unity, $a= 1$.

We also add interaction terms $H_{J'}$ and $H_{D'}$, with neighbors of distance $2a$, which compensate the induced anisotropies originating from the discretization of the corresponding continuum spin model, as was first proposed by Buhrandt and Fritz \cite{Buhrandt13}.
These terms are of the form
\begin{multline}
H_{J'}+ H_{D'}= -J' \sum \limits_{i} {\bf S}_i \cdot ({\bf S}_{i+ 2\hat x}+ {\bf S}_{i+ 2\hat y}+ {\bf S}_{i+ 2\hat z})-\\
-D' \sum \limits_{i}  \left({\bf S}_i \times {\bf S}_{i+ 2\hat x} \cdot \hat x+ {\bf S}_i \times {\bf S}_{i+ 2\hat y} \cdot \hat y+ {\bf S}_i \times {\bf S}_{i+ 2\hat z} \cdot \hat z \right). \label{HD}
\end{multline}
with $J'= -J/16$ and $D'= -D/8$. These additional terms, as we shall demonstrate elsewhere, do not change the character of the phase transition in the system, but shift downward the transition temperature.
The full Hamiltonian for our simulation is $H= H_J+ H_D+ H_{J'}+ H_{D'}$. The length of the helical modulations $\lambda_h$ is governed by the ratio $D/J$, namely, $D/J= \tan (2\pi/\lambda_h)$. For example, the helix with length of 8 lattice sites, $\lambda_h= 8 a$, corresponds approximately to $D/J \simeq 1.0$.

\begin{figure}
\includegraphics[width=1\columnwidth]{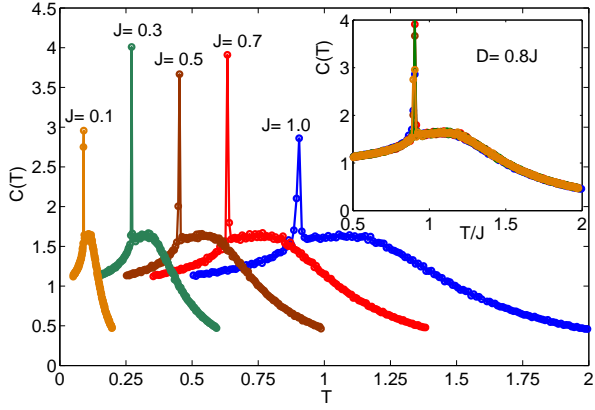}
\caption{Specific heat of the chiral spin system according to classical Monte Carlo calculations at $D/J= 0.8$ and different values of the exchange constant $J$. The inset shows the specific heat as a function of reduced temperature $T/J$.} \label{Fig1}
\end{figure}
\begin{figure}
\includegraphics[width=1\columnwidth]{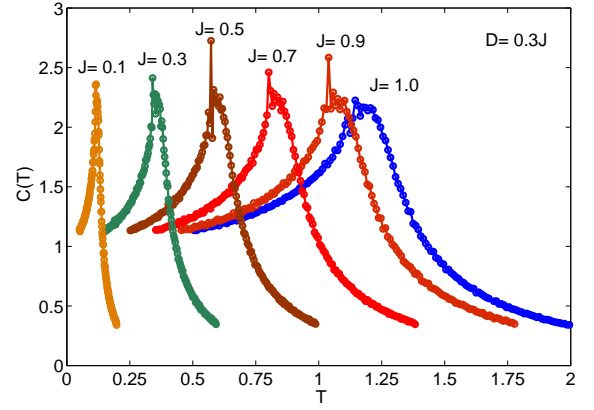}
\caption{Specific heat of the chiral spin system according to classical Monte Carlo calculations at $D/J= 0.3$ and different values of the exchange constant $J$.} \label{Fig2}
\end{figure}
\begin{figure}
\includegraphics[width=1\columnwidth]{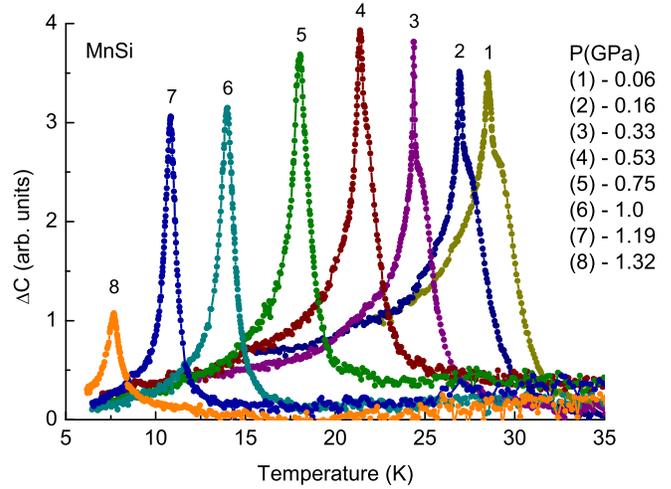}
\caption{Specific heat of the helical magnet MnSi at different pressures \cite{Sidorov14}.} \label{Fig3}
\end{figure}
\begin{figure}
\includegraphics[width=1\columnwidth]{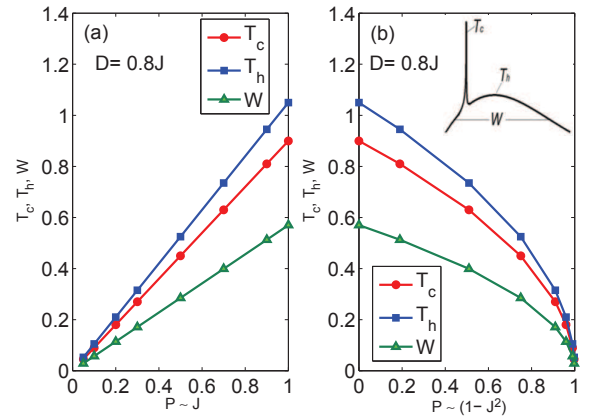}
\caption{Dependence of the phase transition temperature $T_c$, temperature of the «hump» maximum $T_h$, and the «hump» width $W$ (shown in the inset) as  functions of the exchange constant $J$ (a) and $ (1-J^{2})$ (b).} \label{Fig4}
\end{figure}
\begin{figure}
\includegraphics[width=1\columnwidth]{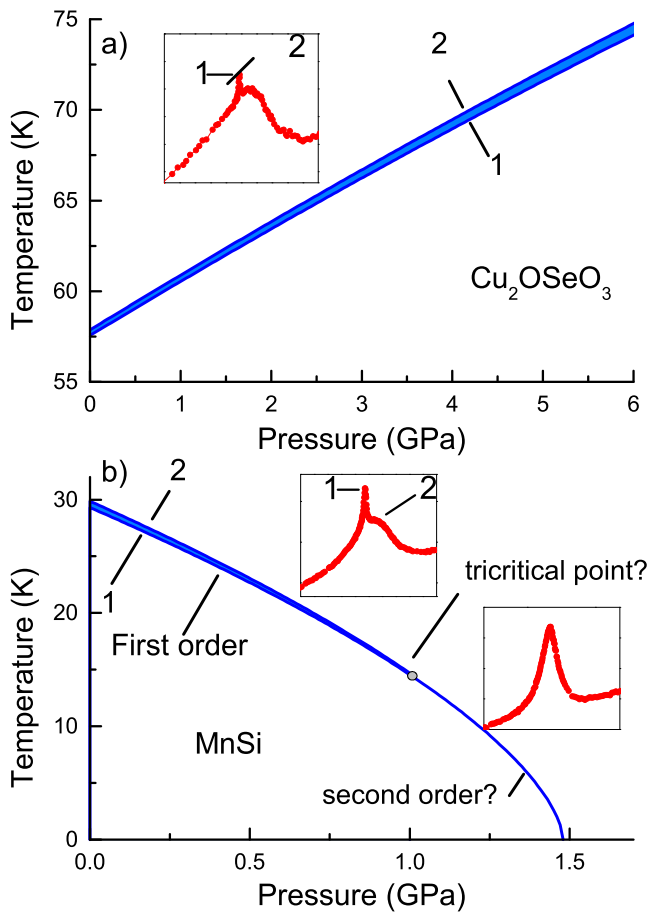}
\caption{Phase diagrams of the helical magnets Cu$_2$OSeO$_3$ and MnSi \cite{Sidorov14}. It is instructive to compare these diagrams with the ones illustrated in Fig. \ref{Fig4}a,b.} \label{Fig5}
\end{figure}

We performed Monte Carlo (MC) simulations with periodic boundary conditions using a standard single-site Metropolis algorithm.
We used a lattice with L= 30 and to check the results we also made simulations for lattice sizes L= 20 and L= 40. To acquire the statistics we usually performed $5 \times 10^5$ MC steps per spin at each temperature.
We start each simulation at temperatures well above $T_c$, and then steadily decreased the temperature by steps of $\Delta T= 10^{-2}J$. We calculate  the energy as a function of temperature $E(T)= \langle H \rangle$ and hence the specific heat $C(T)= (1/N) dE/dT$.  Additional details can be found in a previous paper by the authors \cite{Belemuk17}.

\section{Results and Discussions}
The results of specific heat calculations for the exchange constant $J$ varying from 0.1 to 1.0 at a ratio $D/J$ =0.8 are shown in Fig.\ref{Fig1}. As is known from Ref. ~\cite{Belemuk17}, at this value of $D/J$ and $J$=1, a first order phase transition occurs in this model spin system. Fig. \ref{Fig1} shows the well-known features of the phase transition in the helical magnets: a sharp peak and a rounded "hump" \cite{Stishov11} at different temperatures.  Fluctuating values of the $C(T)$-peak at $T_c$ for different values of $J$ owe to various initial random states used for the simulation at a particular $J$. 
It is instructive to compare the calculations presented in Fig. \ref{Fig1} with the experimental data shown in Fig. \ref{Fig3}.
Curves 1,2 and 3 of Fig. \ref{Fig3}, which are measured at low pressures, show some resemblance with those presented in Fig. \ref{Fig1}.  We see narrowing of the fluctuation anomaly in $C(T)$  with increasing pressure $P$ and decreasing temperature $T_{c}$. However, the sharp peak manifesting a first order phase transition ceases to exist at higher pressures. It gradually moves to low temperatures, simultaneously being transformed into a single maximum, which may indicate a second-order phase transition.

The specific heat calculations at $J$ varying from 0.1 to 1.0 and at the ratio $D/J= 0.3$ are shown in  Fig. \ref{Fig2}. Note that the ratio $D/J \approx 0.3$ is a boundary value, separating regions with and without a first order phase transition ~\cite{Belemuk17}. Again we can see narrowing sets of fluctuation maxima ("humps") with occasional spikes evidencing the proximity of a first order phase transition.

The specific heat curves depicted in Fig. \ref{Fig1}, \ref{Fig2} can be described by a single curve when expressed as a function of a reduced temperature $T/J$ (see the inset in Fig. \ref{Fig1}). Narrowing the fluctuation ("hump") region also follows from scaling, but nevertheless it indicates decreasing the fluctuation population with lowering temperature.   Correspondingly, a dependence of the first order phase transition temperature on the exchange constant is a linear function as shown in Fig. \ref{Fig4}a.
The temperatures of the "hump" maximum, $T_h$, and the «hump» width, $W$, also are presented in Fig. \ref{Fig4}a as functions  of the exchange constant $J$. They are linear functions of $J$ as well. These calculations are applicable to the case of a helical magnet with localized spins, such as found in Cu$_2$OSeO$_3$, where the change of the phase transition temperature at high pressure can be modeled by the corresponding change of $J$.  Experimental data on the pressure dependence of the first order phase transition temperature (curve 1) and the "hump" temperature (curve 2) in Cu$_2$OSeO$_3$ are shown in Fig. \ref{Fig5}a. They demonstrate an almost linear dependence on pressure, that correlates with similar dependencies shown in Fig. \ref{Fig4}a.

A quite different scenario results when the magnetic interaction decreases on compression, as occurs in the itinerant system MnSi.  This situation is modeled in Fig. \ref{Fig4}b using the assumption that pressure can be expressed in the form $P \sim (1-J^{2})$. This form corresponds to the empirical expression $T_{c} \sim (P_{c}-P)^{1/2}$, describing the pressure dependence of $T_{c}$ of a number of the itinerant magnets~\cite{Mohn06}. The corresponding dependencies of $T_c$, $T_h$ and $W$ as functions of $P \sim (1-J^{2})$ imitate quite well the diagram shown in Fig. \ref{Fig5}b.

Returning to Figs. \ref{Fig1}-\ref{Fig3}, we emphasize that, in the our model helical spin system at a condition $D/J \gtrsim 0.3$, a first order phase transition exists at the lowest temperatures, whereas the experimental data show quite a different picture (Fig. \ref{Fig3}). The degradation of the first order phase transition can arise as a natural result of suppression of competing interactions, leading to tricritical phenomena. In this case, we assume an evolution of the ratio $D/J$ along the transition line, as it is evident from Figs. \ref{Fig1}, \ref{Fig2}. However, we cannot exclude the possibility that the observed degradation is caused by a stress inhomogeneity at high pressure. We are not ready to resolve this issue at present.

\section{Conclusion}

We use classical Monte Carlo calculations to model the high-pressure behavior of the phase transition in the helical magnets. Variations of the constants of the exchange interaction $J$ and the Dzyaloshinskii - Moriya interaction $D$ were applied to model the changes in magnetic interactions, which occur in real systems under pressure owing to changing spin-spin distances. Results of the calculations show that an occurrence of the first order phase transition critically depends on the ratio $D/J$.
The phase transition temperature $T_c$ is a linear function of the exchange constant $J$ at fixed value of the ratio $D/J$. High-pressure behavior of the studied spin system depends on a character of evolution of the exchange constant $J$ on compression. The experimental data for helical magnets MnSi and Cu$_2$OSeO$_3$ can be qualitatively reproduced with two suggested modes of evolution of $J$ (see Figs. \ref{Fig4} and \ref{Fig5}). A tricritical behavior can be modeled in the framework of the current spin model if one introduces a variation the ratio $D/J$ along a transition line.

\end{document}